\documentclass[letterpaper]{article}

\usepackage[T1]{fontenc}

\usepackage{geometry}
\usepackage{setspace}
\usepackage{upgreek}
\usepackage[
  backend=biber,
  style=chem-acs,
  articletitle=true,
  autocite=superscript,
  url=false,
  isbn=false
]{biblatex}
\bibliography{nanogap.bib}

\usepackage{graphicx}
\usepackage{float}
\newfloat{scheme}{htbp}{los}
\floatname{scheme}{Scheme}
\floatname{chart}{Chart}
\newfloat{graph}{htbp}{loh}

\usepackage{chemformula} 
\usepackage[version = 4]{mhchem} 

\usepackage{authblk}
\author[1,2]{Dasom Kim}
\author[3$\dagger$]{Dukhyung Lee}
\author[4]{Young-Mi Bahk}
\author[1\footnote{Deceased}]{Dai-Sik Kim}

\affil[1]{Department of Physics, Ulsan National Institute of Science and Technology (UNIST), Ulsan 44919, Republic of Korea}
\affil[2]{Max Planck Institute for the Structure and Dynamics of Matter, Hamburg 22761, Germany}
\affil[3]{College of Physical Sciences and Engineering, Mohammed VI Polytechnic University, 43150, Ben Guerir, Morocco}
\affil[4]{Department of Physics, Incheon National University, Incheon 22012, Republic of Korea}
\date{$^\dagger$Email: dukhyung.lee@um6p.ma}

\title{Mechanical Activation of Terahertz Tunneling \\ in Metallic Nanogaps}

\begin{document}

\maketitle

\begin{abstract}
Metallic nanogaps concentrate terahertz (THz) fields into deep subwavelength volumes and support field-driven electron tunneling when the insulating barrier becomes sufficiently narrow. Here, we demonstrate mechanical control of tunneling-mediated nonlinear THz transmission in a flexible nanogap metasurface. The metasurface consists of Au/PMMA/Au nanogaps fabricated on a polyethylene terephthalate substrate, enabling continuous tuning of the gap geometry through macroscopic bending. In the flat state, the resonant transmission exhibits only a weak dependence on the incident THz field strength. Upon bending, increasing the incident field strength induces pronounced resonance suppression accompanied by saturation of the voltage developed across the nanogaps. This nonlinear response is consistent with the opening of a field-dependent tunneling conduction channel through the mechanically narrowed PMMA barriers. Simmons-model calculations illustrate the strong increase in tunneling current density and the associated dissipative gap response as the local gap width approaches the few-nanometer regime. These results establish mechanical deformation as a macroscopic means of controlling tunneling-mediated THz nonlinearities in flexible metasurfaces.
\end{abstract}

\section*{Keywords}
Terahertz metasurface; Nanogap; Quantum tunneling; Mechanical deformation; Nonlinear terahertz response

\section{Introduction}
Metal–insulator–metal nanogaps confine electromagnetic fields within deep subwavelength volumes and thereby enable strong light--matter interactions\autocite{Seo2009,Savage2012,Kim2018}. In the classical regime, an insulating gap primarily acts as a capacitor that accumulates opposite charges on the opposing metal surfaces\autocite{Bahk2017UltimateNanoslits}. When the barrier width approaches the few-nanometer regime, however, the finite transmission probability of electron wavefunctions permits tunneling across the insulator\autocite{Simmons1963,Simmons1963-2}. The resulting junction can be described as a tunneling conductance in parallel with the gap capacitance. Because the tunneling current depends nonlinearly on both the local electric field and the barrier width, even a small change in the gap geometry can substantially modify the electromagnetic response of the junction\autocite{Guhr2007,Savage2012,Kim2021Topology-ChangingNanotrenches}. The extreme field confinement and electron tunneling interplay and form an essential basis of quantum plasmonics.

Terahertz (THz) radiation provides a particularly useful means of driving and probing ultrafast tunneling. Previous studies have employed THz-driven scanning tunneling microscopy to manipulate transient currents with atomic-scale spatial resolution\autocite{Cocker2013,Jelic2017,Jelic2024} and have incorporated metal–insulator–metal junctions into rectifying antennas for ultrafast electromagnetic energy conversion\autocite{Kang2018,Hemmetter2021,Siday2024}. In parallel, atomic-layer lithography has enabled uniformly patterned metallic nanogaps to be extended over macroscopic areas\autocite{Chen2013}. In such metasurfaces, the collective response of the nanogaps is directly manifested in far-field THz transmission, and intense THz fields can induce a nonlinear transmission response associated with tunneling across the barriers\autocite{Kim2015TerahertzRegime,Bahk2015,Kim2016TunnellingSpectroscopy}. These developments provide a route for controlling quantum transport through the optical response of a macroscopically addressable metasurface.

Active control of tunneling in such macroscopic nanogap metasurfaces nevertheless remains challenging because the current depends critically on the barrier width. Mechanical deformation of flexible substrates offers a direct means of modifying nanogap geometries and has been used to control transmission amplitude, polarization, and resonance frequency\autocite{Pryce2010,Aksu2011,Chen2018}. However, mechanically narrowing the gaps may either activate field-dependent tunneling or establish classical Ohmic metal contacts\autocite{Halbritter2003,KimS2023}. Whether a mechanically deformed nanogap metasurface can retain and controllably modify its tunneling-mediated nonlinearity therefore remains an open question.

Here, we demonstrate mechanical control of nonlinear THz transmission in an Au/poly (methyl methacrylate) (PMMA)/Au nanogap metasurface fabricated on a flexible polyethylene terephthalate (PET) substrate. In the flat state, the transmittance exhibits only a weak dependence on the incident THz field. Upon bending, increasing the incident field suppresses the resonance and produces saturation of the voltage developed across the gaps. Simmons-model calculations illustrate the strong increase in tunneling current density expected as the local PMMA barrier approaches the quantum regime. The tunneling-associated dissipative response is evidenced by the experimentally estimated current density and effective imaginary gap permittivity, supporting a tunneling-mediated origin of the mechanically activated THz nonlinearity.

\section{Results}

\subsection{Flexible Nanogap Metasurface}
Figure 1a illustrates the flexible nanogap metasurface used in this study. The metasurface consists of a periodic array of square-ring Au nanogaps fabricated on a 250-$\upmu$m-thick PET substrate. Each square ring has side lengths of $l_x=l_y=50\,\upmu$m and array periods of $d_x=d_y=100\,\upmu$m. The Au layer has a thickness of 160\,nm. The gap width at the upper opening, defined by the thickness of the PMMA layer, is 30\,nm. The incident THz electric field is polarized perpendicular to two opposing sides of the squares, thereby driving charge accumulation and producing a strongly enhanced local field at these sides. As shown in the cross-sectional scanning electron micrograph in Fig.\,1b, the nanogap has a tapered profile, with a width of approximately 30\,nm at the top and 80\,nm near the bottom. 

The Au/PMMA/Au nanogaps were fabricated through the process illustrated in Fig.\,1c. Au structures were first patterned on the flexible PET substrate, followed by spin coating of a PMMA layer that defined the gap width. A second Au layer was subsequently deposited over the patterned structures. The excess Au was then mechanically removed using adhesive tape, leaving PMMA-filled Au/PMMA/Au nanogaps. This fabrication approach combines nanometer-scale insulating barriers with a macroscopically deformable substrate, allowing the local junction geometry to be controlled through mechanical bending. Details of the nanogap fabrication procedure are provided in Section S1 of the Supporting Information.

\begin{figure}[htb!]
\centering\includegraphics[width=1.\textwidth]{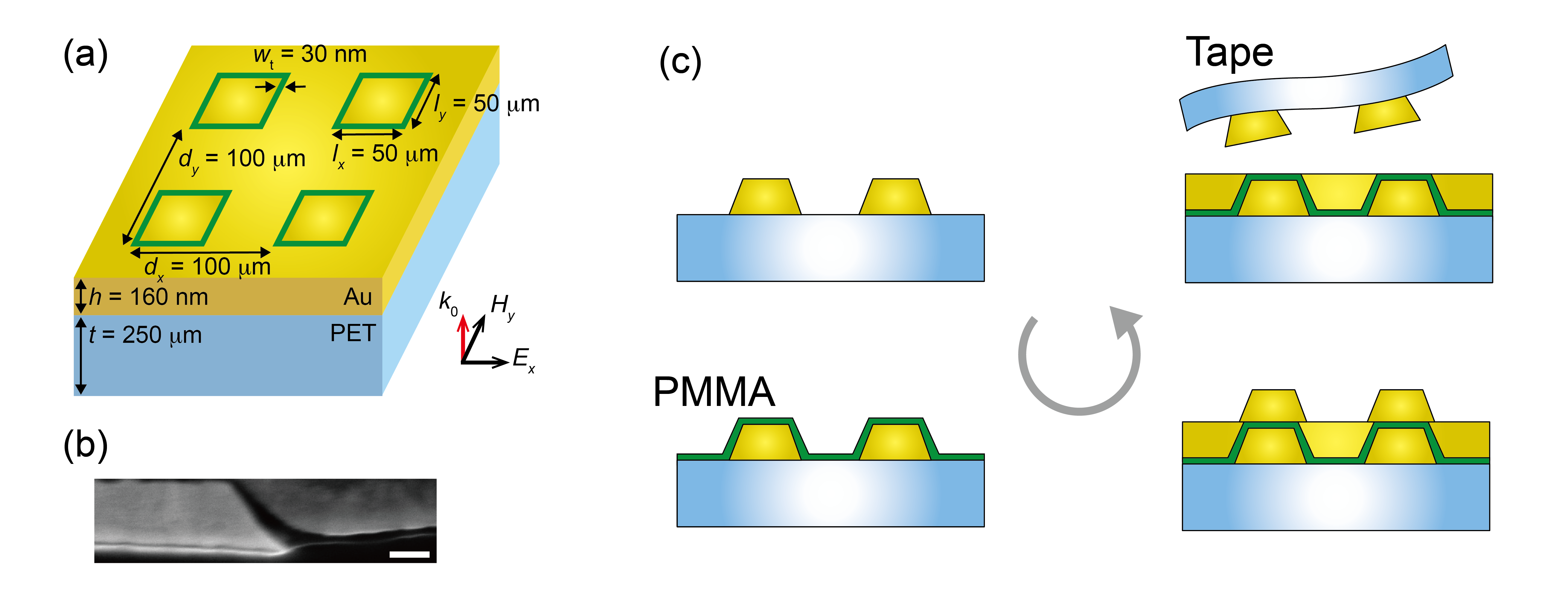}
\caption{Fabrication and geometry of the flexible nanogap metasurface.
(a) Schematic of the metasurface consisting of square-ring Au nanogaps patterned on a 250-$\upmu$m-thick polyethylene terephthalate (PET) substrate. The slot dimensions are \(l_x=l_y=50\,\upmu\mathrm{m}\), the array periods are \(d_x=d_y=100\,\upmu\mathrm{m}\), the gap width is \(w_\text{t}=30\,\mathrm{nm}\), and the Au thickness is \(h=160\,\mathrm{nm}\). The incident THz electric field \(E_x\) is polarized perpendicular to two opposing sides of the squares. (b) The cross-sectional electron micrograph shows the tapered nanogap formed between the adjacent metal surfaces. Scale bar: 100\,nm. (c) Schematic illustration of the fabrication procedure. Prepatterned Au structures on PET are spin-coated with PMMA (green), followed by deposition and planarization of the second Au layer to form PMMA-filled Au/PMMA/Au nanogaps.}
\end{figure}

\subsection{Mechanical Narrowing of the Nanogaps}
Mechanical deformation of the metasurface is illustrated in Fig.\,2a. The initially flat sample has a length of $L_0=25$\,mm. Reducing the end-to-end length to $L$ bends the sample with a corresponding radius $R$, as shown in Fig.\,2b. Bending-induced eccentric compression exerted on the Au layer by the PET substrate deforms the tapered nanogap cross section, preferentially narrowing its upper portion. Details of the bending geometry and the estimation of the bending radius are provided in Section S2 of the Supporting Information.

To examine the geometrical effect of bending, we performed finite-element structural-mechanics simulations of the nanogap cross section. In the simulation, bending was implemented by applying prescribed displacements to both ends of the substrate, thereby mimicking the experimental bending condition. Figure 2c shows the simulated structures before and after bending, together with the corresponding pressure distribution. The simulation predicts that bending preferentially narrows the upper portion of the tapered gap while the lower portion remains relatively wide. As shown in Fig.\,2d, the calculated top-gap width decreases with decreasing bending radius and reaches zero at approximately \(R=16\) mm, indicating nanogap closure. This prediction is obtained from a structural model that neglects the inelastic PMMA within the gap. Despite the substantial local narrowing of the upper gap region predicted by the structural simulation, our experimental results imply that the PMMA within the gap prevents direct metal contacts even under greater bending, as discussed in the following sections. The strongest bending condition used in the experiment corresponds to \(R\approx12\) mm, where current is still dominated by tunneling. For the tunneling calculation below, an effective local barrier width of 1\,nm is assumed as a plausible residual local separation based on the characteristic surface roughness of the opposing Au interfaces.\autocite{Moazzez2013}

The change in gap geometry strongly modifies the THz response in the low-field regime, where the response remains linear with the incident field strength. As shown in Fig.\,2e, the flat metasurface exhibits a resonant transmission peak near 0.67\,THz, originating from capacitive charging across the nanogaps. When the sample is bent, the resonant transmission is strongly suppressed. This change reflects the reduced separation of the opposing metal surfaces and the resulting modification of the gap impedance. The mechanically induced suppression of the low-field transmission establishes that macroscopic bending provides effective control over the nanoscale junction geometry.

\begin{figure}[htb!]
\centering\includegraphics[width=0.55\textwidth]{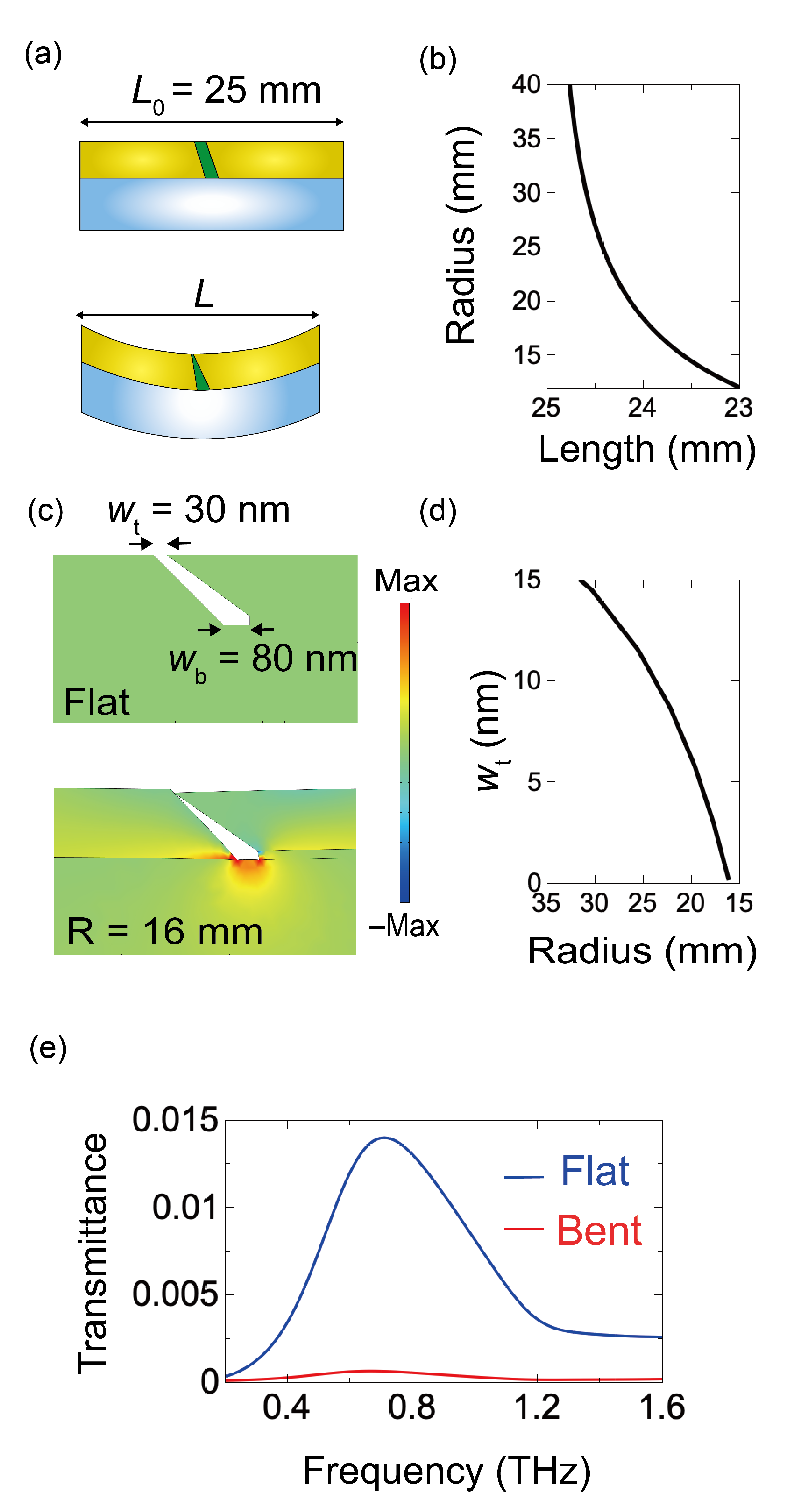}
\caption{Mechanical narrowing and low-field THz response of the flexible nanogap metasurface.
(a) Cross-sectional schematics of the metasurface in the flat and bent states. \(L_0=25\,\mathrm{mm}\) and \(L\) denote the initial and bent sample lengths, respectively. (b) Bending radius calculated as a function of \(L\). (c) Finite-element structural-mechanics simulations of the nanogap cross section before and after bending, respectively. The color map shows the pressure distribution near the gap. PMMA is omitted from the simulations. (d) Upper-end gap width as a function of bending radius calculated from the simulation. (e) Low-field THz transmission spectra of the metasurface in the flat (blue) and bent ($R=12$\,mm, red) states, showing strong suppression of the resonant transmission upon bending.}
\end{figure}

\subsection{Tunneling across the Narrowed PMMA Barrier}
The electrical response of the nanogap can be modeled as a gap capacitance in parallel with a tunneling conduction channel, as illustrated in Fig.\,3a. For a wide insulating barrier, the junction is predominantly capacitive because the electron transmission probability through the PMMA layer is negligible (Fig.\,3b, top). As the local barrier width approaches the nanometer regime upon bending, the electron wavefunction penetrates through the barrier and produces a finite tunneling current. The voltage across the nanogap additionally tilts and lowers the effective barrier, causing the tunneling probability to increase nonlinearly with the local electric field (Fig.\,3b, bottom).

We evaluated this field-dependent response using the Simmons tunneling model. The locally narrowed portion of the junction was approximated as two parallel Au surfaces separated by a PMMA barrier of width $w$, from which the local tunneling current density can be determined. Figure 3c presents the calculated current density as a function of gap voltage for barrier widths ranging from 1 to 20\,nm. The current density exhibits a nonlinear dependence on both the barrier width and the applied voltage. For gap widths of 10–20\,nm, the calculated current density remains negligible over most of the examined voltage range. In contrast, reducing the gap to 1–3\,nm produces an increase of several orders of magnitude in the tunneling current density. These calculations demonstrate that narrowing of the nanogap can activate a substantial conduction channel. Details of the Simmons-model calculation of the tunneling current are provided in Section S3 of the Supporting Information.

\begin{figure}[tb!]
\centering\includegraphics[width=0.5\textwidth]{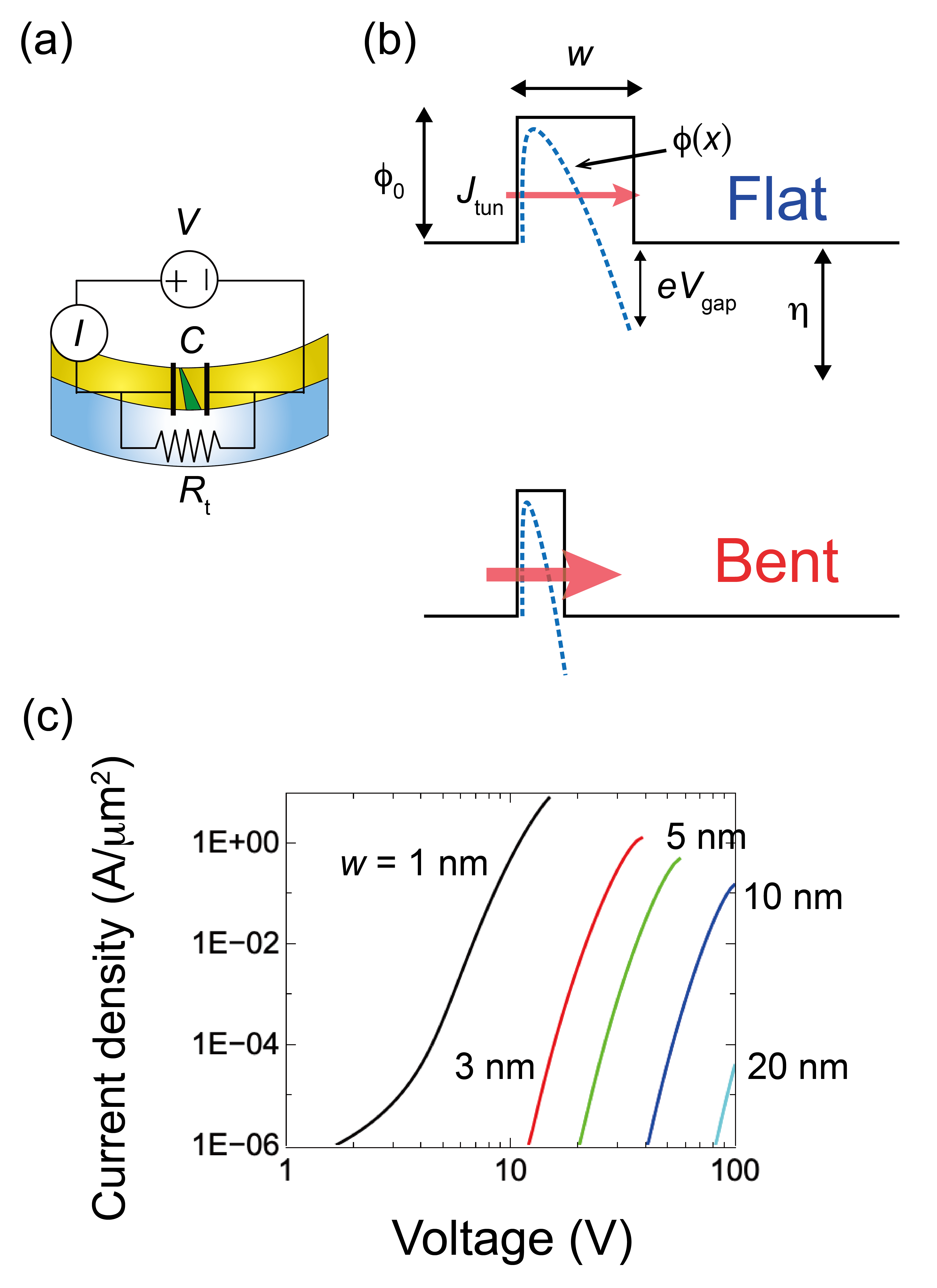}
\caption{Calculated tunneling response of a mechanically narrowed Au/PMMA/Au nanogap.
(a) Equivalent circuit representation of the bent nanogap, in which the gap capacitance is accompanied by a tunneling conduction channel across the narrowed PMMA barrier. (b) Schematic potential-energy profiles of wide and narrow barriers under an applied gap voltage (dashed lines) and zero bias (solid lines) . Here, $\phi_0 = 4.8$\,eV is the zero-bias barrier height, $\phi(x)$ is the voltage-dependent barrier profile, $eV_{\mathrm{gap}}$ is the energy drop across the junction, $\eta$ is the Au Fermi energy, and $J_{\mathrm{tun}}$ is the tunneling current density. Barrier narrowing and field-induced tilting increase the electron tunneling probability. (c) Tunneling current density calculated as a function of gap voltage using the Simmons model for parallel Au/PMMA/Au junctions with gap widths of 1, 3, 5, 10, and 20\,nm. The tunneling current increases by several orders of magnitude as the gap width decreases toward the nanometer regime.}
\end{figure}

\subsection{Mechanically Activated Nonlinear THz Transmission}
We investigated the nonlinear response of the metasurface using intense single-cycle THz pulses, as illustrated in Fig.\,4a. The THz electric field was polarized perpendicular to the mechanically narrowed sides of the nanogaps. Measurements were performed using three different THz pulse intensities, corresponding to incident peak electric fields of 75, 108, and 150\,kV/cm, for both the flat and strongly bent configurations ($R = 12$\,mm). Details of the nonlinear THz spectroscopy measurements are provided in Section S4 of the Supporting Information.

The role of the incident THz pulse can be understood in terms of an equivalent current-driven junction model\autocite{Kim2016TunnellingSpectroscopy,Park2026}. For normally incident radiation polarized perpendicular to the nanogap, the magnetic field induces a surface current on the optically thick Au layer. In the perfect-conductor approximation, the driving surface-current density is
\begin{equation}
\mathbf{K}_0(t)
=
\hat{\mathbf{n}}
\times
2\mathbf{H}_{\mathrm{inc}}(t),
\label{eq:surface_current}
\end{equation}
where $\hat{\mathbf{n}}$ is the surface-normal unit vector and $\mathbf{H}_{\mathrm{inc}}(t)$ is the incident THz magnetic field. Because this surface current is determined primarily by the incident THz pulse, the pulse acts as an externally prescribed transient current source for the junction (Fig.\,4a right). The gap voltage $V_{\mathrm{gap}}(t)$ evolves dynamically from the competition between capacitive charge accumulation and tunneling discharge: the driving current accumulates opposite charges on the facing Au surfaces, increasing $V_{\mathrm{gap}}$, while the resulting voltage drives electron tunneling $J_{\mathrm{tun}}(V)$ described by the Simmons model, which discharges the junction and reduces $V_{\mathrm{gap}}$.


In the THz experiment, $V_{\mathrm{gap}}(t)$ was estimated from the transmitted THz waveform using an area-averaging approximation based on the Kirchhoff integral formalism. Figure 4b shows the voltage waveforms developed across the nanogaps. The gap field is much larger than the measured far-field transmission because, at the exit surface of the metasurface, the transmitted field is confined to the small fraction of the unit-cell area occupied by the nanogaps. The gap voltage was estimated as
\begin{equation}
V_{\mathrm{gap}}(t) = E_\text{gap}(t)w_\text{t} = \frac{w_\text{t}}{\beta}\frac{4n_{\mathrm{PET}}}{(1+n_{\mathrm{PET}})^2} \frac{E_{\mathrm{sam}}(t)} {E_{\mathrm{PET}}^{\mathrm{pk}}}E_0,
\end{equation}
where \(E_{\mathrm{gap}}(t)\) is the THz field in the nanogap, \(E_{\mathrm{sam}}(t)\) is the THz field transmitted through the sample, \(E_{\mathrm{PET}}^{\mathrm{pk}}\) is the peak field transmitted through the bare PET reference, \(E_0\) is the incident peak field, and \(n_{\mathrm{PET}}=1.78\). The nanogap coverage ratio is \(\beta=2l_yw_\text{t}/(p_xp_y)\), where the factor of 2 accounts for the two sides of each square-ring nanogap oriented perpendicular to the incident electric field. Because \(w_\text{t}/\beta=p_xp_y/(2l_y)\), the estimated gap voltage is independent of the gap width. Details of the estimation of the voltage across the nanogaps from the measured far-field THz transmission are provided in Section S5 of the Supporting Information.

In the flat state, the gap voltage clearly increases with the incident THz field strength, while retaining a similar temporal waveform as shown in Fig.\,4b. In contrast, the bent state exhibits much smaller gap voltages and a substantially weaker increase with incident field strength. The peak gap voltages are summarized in Fig.\,4c. For the flat metasurface, the peak voltage increases almost linearly with incident field, reaching approximately 85\,V at the incident peak field of 150\,kV/cm. For the bent metasurface, the voltage increases only from approximately 10 to 12\,V over the same incident field range, indicating pronounced saturation.

This contrast can be understood from the field-dependent impedance of the nanogap. In the flat state, the relatively wide PMMA barrier suppresses tunneling, allowing charges to accumulate capacitively at the opposing metal surfaces. This charge accumulation without conduction channel produces a gap voltage that scales approximately with the incident field strength. In the bent state, however, the mechanically narrowed barrier permits electron tunneling. The tunneling current partially discharges the accumulated surface charges and limits further enhancement of the electric field across the gap, resulting in saturation of the gap voltage.

To estimate the local tunneling response, the experimentally determined peak gap voltages were applied to the Simmons model. For this comparison, the flat and bent junctions were approximated as parallel Au/PMMA/Au gaps with widths of 30 and 1\,nm, respectively. The calculated tunneling current densities are shown in Fig.\,4d. The current density remains negligible for the flat configuration because of its substantially wide barrier. For the bent configuration, it increases from approximately 0.8 to 2.2\,A/$\upmu$m$^2$ over the measured incident-field range. The corresponding dissipative response can be represented by the imaginary part of effective dielectric constant,
\begin{equation}
\operatorname{Im}\{\varepsilon_{\mathrm{gap}}\} =
\frac{J_{\mathrm{tun}}} {\varepsilon_0\omega E_{\mathrm{gap}}^{\mathrm{pk}}},
\end{equation}
where $E_{\mathrm{gap}}^{\mathrm{pk}}$ is the peak THz field in the nanogap, $\omega$ is the angular frequency, $\epsilon_0$ is the vacuum permittivity, and \(J_{\mathrm{tun}}\) is the calculated tunneling current density. Figure 4e shows the resulting \(\operatorname{Im}\{\varepsilon_{\mathrm{gap}}\}\) at the resonance frequency of 0.67 THz. Details of the conversion of the tunneling current into the effective dielectric response of the PMMA barrier are provided in Section S6 of the Supporting Information. The imaginary part of effective dielectric constant increases with incident field in the bent state, indicating an increasing dissipative contribution from field-driven tunneling. The simultaneous saturation of the gap voltage and increase in the calculated dissipative response are consistent with the activation of a nonlinear tunneling channel, rather than the formation of linear Ohmic contacts.

\begin{figure}[tb!]
\centering\includegraphics[width=1\textwidth]{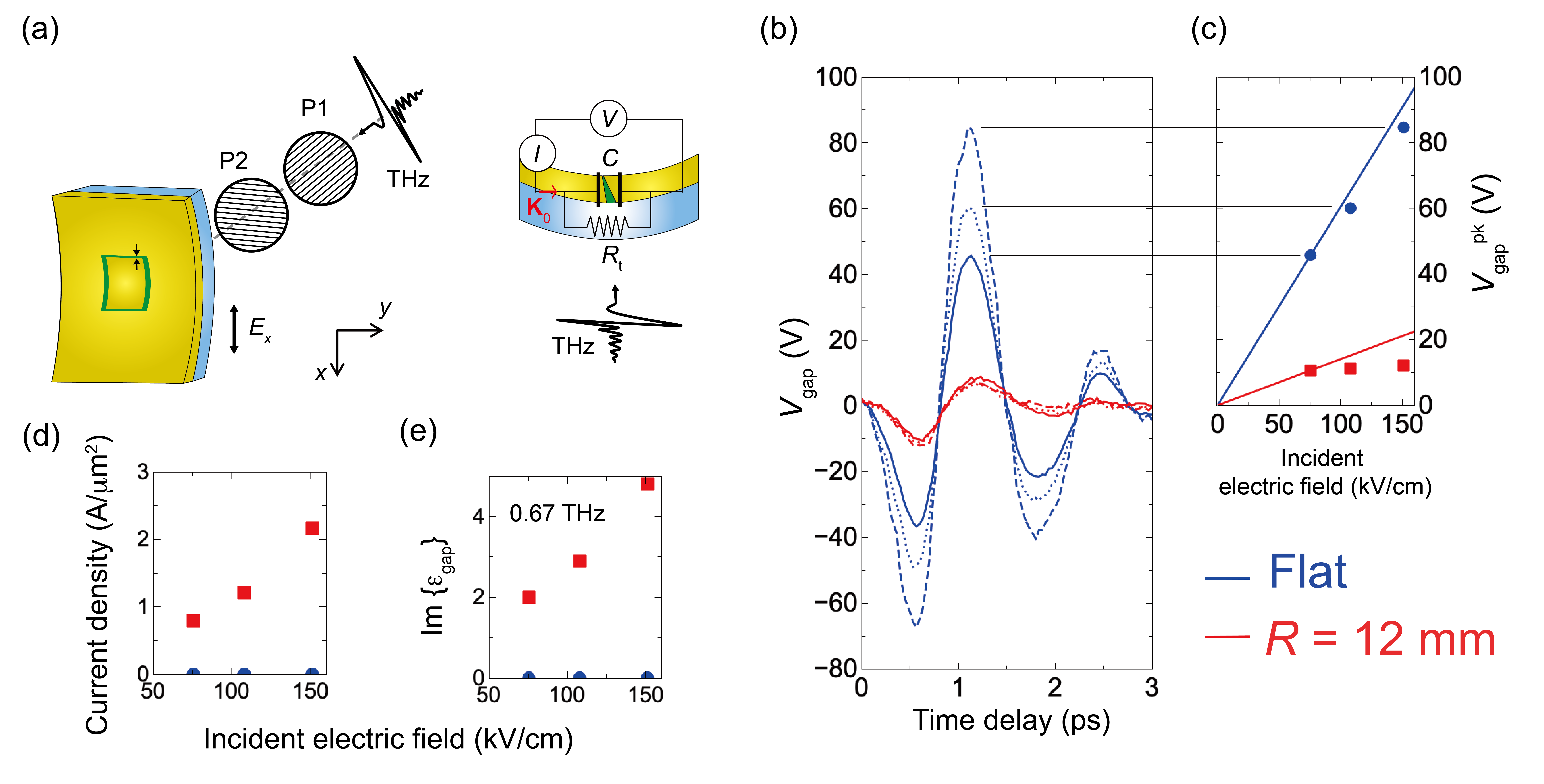}
\caption{Gap voltage under high-field THz excitation and estimated tunneling response.
(a) Schematic of the high-field THz transmission measurement performed on the mechanically bent nanogap metasurface. The incident THz electric field is polarized perpendicular to the narrowed sides of the nanogaps. (b) Temporal waveforms of the voltage developed across the nanogaps in the flat (blue) and bent ($R=12$\,mm, red) states at 75 (solid line), 108 (dotted line), and 150\,kV/cm (dashed line). (c) Peak gap voltage as a function of the incident electric field strength for the flat and bent states. In the flat state, the gap voltage increases approximately linearly with the incident field, whereas it tends to saturate in the bent state. (d) Tunneling current density calculated by applying the experimentally determined peak gap voltages to the Simmons model. The locally narrowed region in the bent state was approximated as a parallel Au/PMMA/Au junction with a gap width of 1\,nm. (e) Corresponding effective imaginary permittivity of the gap at 0.67\,THz. The increasing dissipative response in the bent state is consistent with the activation of field-driven tunneling across the mechanically narrowed barrier.}
\end{figure}

The mechanical evolution of the field-dependent response becomes more evident in the frequency domain. Figure 5a shows the transmission spectra in the flat and strongly bent ($R=12$\,mm) configurations, respectively. In the flat state, the resonance near 0.67\,THz exhibits only a modest reduction with increasing incident field. Under strong bending, the resonant transmission is progressively suppressed and approaches the nonresonant background.

Figure 5b shows the resonant transmittance for each configuration normalized by its respective value at 75\,kV/cm. At the highest incident field, the normalized resonant transmittance remains approximately 0.9 in the flat state, decreases to approximately 0.7 under intermediate bending, and reaches approximately 0.2 under the strongest deformation. This systematic increase in nonlinear suppression with bending curvature demonstrates mechanical control of the field-dependent response and is consistent with the progressive activation of tunneling-assisted dissipation in locally narrowed PMMA barriers.


\begin{figure}[tb!]
\centering\includegraphics[width=0.4\textwidth]{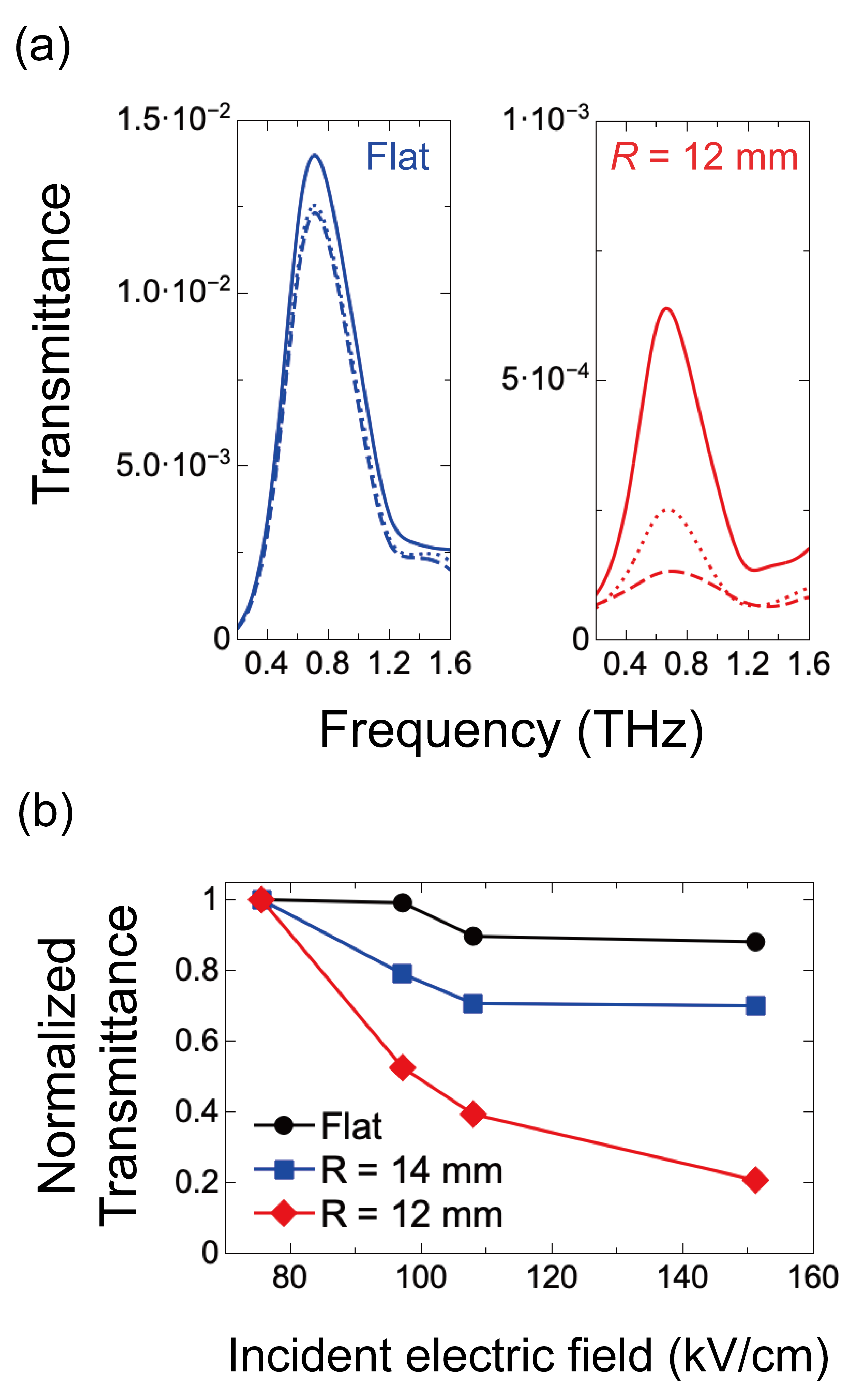}
\caption{Mechanically activated nonlinear THz transmission. (a) THz transmittance spectra of the nanogap metasurface in the flat and bent states, respectively, obtained by Fourier transformation of the time-domain waveforms in Fig.\,4 and normalized to the transmittance through a bare PET substrate. The solid, dotted, and dashed curves correspond to incident field strengths of 75, 108, and 150\,kV/cm, respectively. (b) Normalized resonant transmittance as a function of incident field strength for the flat, intermediately bent ($R=14$\,mm), and strongly bent ($R=12$\,mm) configurations, including an additional measurement at 97\,kV/cm. For each bending condition, the transmittance is normalized to its corresponding value at 75\,kV/cm, $T_{\mathrm{norm}}^{\mathrm{res}}(E_0) =T^{\mathrm{res}}(E_0)/T^{\mathrm{res}}(75\,\mathrm{kV/cm})$.}
\end{figure}

\section{Discussion}
The results demonstrate that mechanical deformation controls not only the low-field linear resonance of the nanogap metasurface but also its nonlinear response to intense THz fields. Bending reduces the minimum width of the tapered PMMA barrier from the tens of nanometers toward the $\sim$1-nm scale, as expected from the structural-mechanics simulations. This geometric change strongly suppresses the low-field transmission and, more importantly, produces a pronounced field-dependent response that is absent in the flat state. The nonlinear response therefore emerges only when mechanical deformation brings the opposing metal surfaces sufficiently close to enable an additional conduction channel across the gap.

The contrasting behaviors of the flat and bent metasurfaces can be understood in terms of the competition between capacitive charging and tunneling-assisted discharge. In the flat state, the wide PMMA barrier prevents appreciable electron transport, and the junction behaves predominantly as a capacitor. Charges accumulated at the opposing metal surfaces generate a strongly enhanced local field, resulting in a gap voltage that increases approximately linearly with the incident THz field. In the strongly bent state, the upper portion of the tapered gap is expected to narrow substantially. Although direct determination of the residual barrier thickness is difficult, the Kirchhoff formalism allows the gap voltage to be determined independently of the gap width. We use 1\,nm as a representative roughness-limited local barrier width to capture the characteristic tunneling response. The resulting increase in tunneling probability allows electrons to cross the barrier and partially discharge the accumulated surface charges. Consequently, the gap voltage no longer follows the incident field linearly but instead exhibits saturation, accompanied by a progressive suppression of the transmission.

The observed nonlinearity is distinct from the response expected from a static metallic contact. A permanent Ohmic connection would suppress capacitive charging across the entire range of incident field strengths and would therefore primarily produce a field-independent reduction of the resonance\autocite{Kim2021Topology-ChangingNanotrenches}. In contrast, the bent metasurface retains a measurable resonance at the lowest incident field, which is further suppressed as the field increases. This field-dependent behavior is consistent with tunneling through a finite PMMA barrier, for which the current depends nonlinearly on the gap voltage. The Simmons calculations support this interpretation by showing that the current density increases by several orders of magnitude as the barrier width decreases from tens of nanometers into the nanometer-scale tunneling regime.

A notable feature of this platform is the coupling between a macroscopic mechanical 
parameter to an ultrafast nanoscale electrical response. The gap geometry is controlled through millimeter-scale bending, whereas the resulting electron transport occurs across a nanometer-scale barrier within the duration of a THz pulse. This flexible platform therefore connects mechanical deformation to an ultrafast tunneling-mediated optical response. Unlike fixed nanogap structures, in which the tunneling regime is set during fabrication, the present metasurface allows the nonlinear response to be activated mechanically after fabrication.

More generally, these results suggest a route toward mechanically reconfigurable nonlinear THz components, beyond the linear regime\autocite{Pitchappa2016,Manjappa2018,Prakash2024}. The far-field transmission provides an optical readout of the local junction response across a macroscopically extended metasurface, without requiring electrical contacts to individual nanogaps. Further control over the barrier profile, deformation geometry, and mechanical stability could enable continuous tuning between capacitive, tunneling-assisted, and contact regimes. Such control may facilitate the development of flexible THz modulators and ultrafast nonlinear switches, while providing a mechanically tunable platform for investigating ultrafast transport in nanoscale junctions.

\section{Conclusion}
In conclusion, we demonstrated mechanical control of tunneling-mediated nonlinear THz transmission in a flexible Au/PMMA/Au nanogap metasurface. Structural-mechanics simulations show that bending narrows the upper portion of the tapered PMMA barrier toward the nanometer regime. Whereas the flat metasurface exhibits an approximately linear increase in gap voltage with incident field, the bent metasurface shows voltage saturation and progressive suppression of the transmission. Simmons-model calculations attribute this behavior to a rapidly increasing tunneling current and the associated dissipative gap response. The field-dependent response of the bent structure is consistent with tunneling through a finite PMMA barrier rather than a static metallic contact. Our results establish macroscopic mechanical deformation as a means of controlling nanometer-scale ultrafast transport and nonlinear THz response in flexible metasurfaces.

\clearpage

\section{Associated Content}
\subsection{Supporting Information}
Detailed fabrication procedure for the flexible Au/PMMA/Au nanogap metasurface; estimation of the bending radius and curvature uniformity across the THz beam; Simmons-model calculation of the tunneling current; experimental setup and calibration for high-field THz measurements; estimation of the voltage across the nanogaps from the far-field THz transmission; conversion of the tunneling current into the effective dielectric response of the PMMA barrier.

\section{Author Information}
\subsection{Author Contributions}
D.K.\ fabricated the sample, performed the experiments and analysis, and wrote
the manuscript based on discussions with Y.-M.B., D.L., and D.-S.K. The experimental concept was conceived by D.K.\ and D.-S.K. D.L.\ performed COMSOL simulation and wrote the manuscript. All authors reviewed and contributed to the final manuscript.

\subsection{Funding}
This work was supported by the National Research Foundation of Korea (NRF) grant funded by the Korean government (NRF-2015R1A3A2031768, NRF-2022R1I1A1A01073838). 

\subsection{Notes}
The authors declare no competing interest.

\section*{Acknowledgements}
The authors thank Bamadev Das, Hyeong Seok Yun, and Dohee Lee for useful discussions.

\printbibliography

@article{Seo2009,
   author = {Seo, M. A. and Park, H. R. and Koo, S. M. and Park, D. J. and Kang, J. H. and Suwal, O. K. and Choi, S. S. and Planken, P. C. M. and Park, G. S. and Park, N. K. and Park, Q. H. and Kim, Dai-Sik},
   title = {Terahertz field enhancement by a metallic nano slit operating beyond the skin-depth limit},
   journal = {Nat. Photon.},
   volume = {3},
   number = {3},
   pages = {152-156},
   ISSN = {1749-4893},
   DOI = {10.1038/nphoton.2009.22},
   url = {https://doi.org/10.1038/nphoton.2009.22},
   year = {2009},
   type = {Journal Article}
}

@article{Kim2018,
   author = {Kim, Dasom and Jeong, Jeeyoon and Choi, Geunchang and Bahk, Young-Mi and Kang, Taehee and Lee, Dukhyung and Thusa, Bidhek and Kim, Dai-Sik},
   title = {Giant Field Enhancements in Ultrathin Nanoslots above 1 Terahertz},
   journal = {ACS Photonics},
   volume = {5},
   number = {5},
   pages = {1885-1890},
   DOI = {10.1021/acsphotonics.8b00151},
   url = {https://doi.org/10.1021/acsphotonics.8b00151},
   year = {2018},
   type = {Journal Article}
}

@article{Kim2015TerahertzRegime,
    title = {Terahertz Quantum Plasmonics of Nanoslot Antennas in Nonlinear Regime},
    year = {2015},
    journal = {Nano Lett.},
    author = {Kim, Joon-Yeon and Kang, Bong Joo and Park, Joohyun and Bahk, Young-Mi and Kim, Won Tae and Rhie, Jiyeah and Jeon, Hyeongtag and Rotermund, Fabian and Kim, Dai-Sik},
    number = {10},
    pages = {6683--6688},
    volume = {15},
    publisher = {American Chemical Society},
    url = {http://dx.doi.org/10.1021/acs.nanolett.5b02505},
    doi = {10.1021/acs.nanolett.5b02505},
    issn = {1530-6984}
}

@article{Kim2016TunnellingSpectroscopy,
    title = {{Tunnelling current-voltage characteristics of Angstrom gaps measured with terahertz time-domain spectroscopy}},
    year = {2016},
    journal = {Sci. Rep.},
    author = {Kim, Joon-Yeon and Kang, Bong Joo and Bahk, Young-Mi and Kim, Yong Seung and Park, Joohyun and Kim, Won Tae and Rhie, Jiyeah and Han, Sanghoon and Jeon, Hyeongtag and Park, Cheol-Hwan and Rotermund, Fabian and Kim, Dai-Sik},
    pages = {29103},
    volume = {6},
    url = {https://www.ncbi.nlm.nih.gov/pmc/articles/PMC4928118/pdf/srep29103.pdf},
    issn = {2045-2322}
}

@article{Bahk2017UltimateNanoslits,
    title = {{Ultimate terahertz field enhancement of single nanoslits}},
    year = {2017},
    journal = {Phys. Rev. B},
    author = {Bahk, Young-Mi and Han, Sanghoon and Rhie, Jiyeah and Park, Joohyun and Jeon, Hyeongtag and Park, Namkyoo and Kim, Dai-Sik},
    number = {7},
    pages = {75424},
    volume = {95},
    publisher = {American Physical Society},
    url = {https://link.aps.org/doi/10.1103/PhysRevB.95.075424}
}

@article{Kim2021Topology-ChangingNanotrenches,
    title = {{Topology-changing broadband metamaterials enabled by closable nanotrenches}},
    year = {2021},
    journal = {Nano Lett.},
    author = {Kim, Dasom and Yun, Hyeong Seok and Das, Bamadev and Rhie, Jiyeah and Vasa, Parinda and Kim, Young-Il and Choa, Sung-Hoon and Park, Namkyoo and Lee, Dukhyung and Bahk, Young-Mi and Kim, Dai-Sik},
    number = {10},
    pages = {4202--4208},
    volume = {21},
    publisher = {American Chemical Society},
    url = {https://doi.org/10.1021/acs.nanolett.1c00025 https://pubs.acs.org/doi/pdf/10.1021/acs.nanolett.1c00025},
    doi = {10.1021/acs.nanolett.1c00025},
    issn = {1530-6984}
}

@article{Simmons1963,
   author = {Simmons, John G.},
   title = {Generalized Formula for the Electric Tunnel Effect between Similar Electrodes Separated by a Thin Insulating Film},
   journal = {Journal of Applied Physics},
   volume = {34},
   number = {6},
   pages = {1793-1803},
   DOI = {10.1063/1.1702682},
   url = {https://aip.scitation.org/doi/abs/10.1063/1.1702682
https://aip.scitation.org/doi/pdf/10.1063/1.1702682},
   year = {1963},
   type = {Journal Article}
}

@article{Simmons1963-2,
   author = {Simmons, John G.},
   title = {Electric Tunnel Effect between Dissimilar Electrodes Separated by a Thin Insulating Film},
   journal = {Journal of Applied Physics},
   volume = {34},
   number = {9},
   pages = {2581-2590},
   DOI = {10.1063/1.1729774},
   url = {https://aip.scitation.org/doi/abs/10.1063/1.1729774
https://aip.scitation.org/doi/pdf/10.1063/1.1729774},
   year = {1963},
   type = {Journal Article}
}

@article{Guhr2007,
   author = {Guhr, D. C. and Rettinger, D. and Boneberg, J. and Erbe, A. and Leiderer, P. and Scheer, E.},
   title = {Influence of Laser Light on Electronic Transport through Atomic-Size Contacts},
   journal = {Physical Review Letters},
   volume = {99},
   number = {8},
   pages = {086801},
   DOI = {10.1103/PhysRevLett.99.086801},
   url = {https://link.aps.org/doi/10.1103/PhysRevLett.99.086801
https://journals.aps.org/prl/pdf/10.1103/PhysRevLett.99.086801},
   year = {2007},
   type = {Journal Article}
}

@article{Pryce2010,
   author = {Pryce, Imogen M. and Aydin, Koray and Kelaita, Yousif A. and Briggs, Ryan M. and Atwater, Harry A.},
   title = {Highly Strained Compliant Optical Metamaterials with Large Frequency Tunability},
   journal = {Nano Letters},
   volume = {10},
   number = {10},
   pages = {4222-4227},
   ISSN = {1530-6984},
   DOI = {10.1021/nl102684x},
   url = {https://doi.org/10.1021/nl102684x
https://pubs.acs.org/doi/pdf/10.1021/nl102684x},
   year = {2010},
   type = {Journal Article}
}

@article{Aksu2011,
   author = {Aksu, Serap and Huang, Min and Artar, Alp and Yanik, Ahmet A. and Selvarasah, Selvapraba and Dokmeci, Mehmet R. and Altug, Hatice},
   title = {Flexible Plasmonics on Unconventional and Nonplanar Substrates},
   journal = {Advanced Materials},
   volume = {23},
   number = {38},
   pages = {4422-4430},
   ISSN = {0935-9648},
   DOI = {10.1002/adma.201102430},
   url = {https://onlinelibrary.wiley.com/doi/abs/10.1002/adma.201102430
https://onlinelibrary.wiley.com/doi/pdfdirect/10.1002/adma.201102430?download=true},
   year = {2011},
   type = {Journal Article}
}

@article{Savage2012,
   author = {Savage, Kevin J. and Hawkeye, Matthew M. and Esteban, Rubén and Borisov, Andrei G. and Aizpurua, Javier and Baumberg, Jeremy J.},
   title = {Revealing the quantum regime in tunnelling plasmonics},
   journal = {Nature},
   volume = {491},
   number = {7425},
   pages = {574-577},
   ISSN = {1476-4687},
   DOI = {10.1038/nature11653},
   url = {https://doi.org/10.1038/nature11653
https://www.nature.com/articles/nature11653.pdf},
   year = {2012},
   type = {Journal Article}
}

@article{Chen2013,
   author = {Chen, Xiaoshu and Park, Hyeong-Ryeol and Pelton, Matthew and Piao, Xianji and Lindquist, Nathan C and Im, Hyungsoon and Kim, Yun Jung and Ahn, Jae Sung and Ahn, Kwang Jun and Park, Namkyoo and Kim, Dai-Sik and Oh, Sang-Hyun},
   title = {Atomic layer lithography of wafer-scale nanogap arrays for extreme confinement of electromagnetic waves},
   journal = {Nature communications},
   volume = {4},
   pages = {2361},
   ISSN = {2041-1723},
   url = {https://www.nature.com/articles/ncomms3361.pdf},
   year = {2013},
   type = {Journal Article}
}

@article{Cocker2013,
   author = {Cocker, Tyler L. and Jelic, Vedran and Gupta, Manisha and Molesky, Sean J. and Burgess, Jacob A. J. and Reyes, Glenda De Los and Titova, Lyubov V. and Tsui, Ying Y. and Freeman, Mark R. and Hegmann, Frank A.},
   title = {An ultrafast terahertz scanning tunnelling microscope},
   journal = {Nature Photonics},
   volume = {7},
   number = {8},
   pages = {620-625},
   ISSN = {1749-4893},
   DOI = {10.1038/nphoton.2013.151},
   url = {https://doi.org/10.1038/nphoton.2013.151},
   year = {2013},
   type = {Journal Article}
}

@article{Bahk2015,
   author = {Bahk, Young-Mi and Kang, Bong Joo and Kim, Yong Seung and Kim, Joon-Yeon and Kim, Won Tae and Kim, Tae Yun and Kang, Taehee and Rhie, Jiyeah and Han, Sanghoon and Park, Cheol-Hwan and Rotermund, Fabian and Kim, Dai-Sik},
   title = {Electromagnetic saturation of angstrom-sized quantum barriers at terahertz frequencies},
   journal = {Physical review letters},
   volume = {115},
   number = {12},
   pages = {125501},
   url = {https://journals.aps.org/prl/abstract/10.1103/PhysRevLett.115.125501
https://journals.aps.org/prl/pdf/10.1103/PhysRevLett.115.125501},
   year = {2015},
   type = {Journal Article}
}

@article{Chen2018,
   author = {Chen, Wenxiang and Liu, Wenjing and Jiang, Yijie and Zhang, Mingliang and Song, Naixin and Greybush, Nicholas J. and Guo, Jiacen and Estep, Anna K. and Turner, Kevin T. and Agarwal, Ritesh and Kagan, Cherie R.},
   title = {Ultrasensitive, Mechanically Responsive Optical Metasurfaces via Strain Amplification},
   journal = {ACS Nano},
   volume = {12},
   number = {11},
   pages = {10683-10692},
   ISSN = {1936-0851},
   DOI = {10.1021/acsnano.8b04889},
   url = {https://doi.org/10.1021/acsnano.8b04889
https://pubs.acs.org/doi/pdf/10.1021/acsnano.8b04889},
   year = {2018},
   type = {Journal Article}
}

@article{Kang2018,
   author = {Kang, Taehee and Kim, R. H. Joon-Yeon and Choi, Geunchang and Lee, Jaiu and Park, Hyunwoo and Jeon, Hyeongtag and Park, Cheol-Hwan and Kim, Dai-Sik},
   title = {Terahertz rectification in ring-shaped quantum barriers},
   journal = {Nature Communications},
   volume = {9},
   number = {1},
   pages = {4914},
   ISSN = {2041-1723},
   DOI = {10.1038/s41467-018-07365-w},
   url = {https://doi.org/10.1038/s41467-018-07365-w
https://www.ncbi.nlm.nih.gov/pmc/articles/PMC6249216/pdf/41467_2018_Article_7365.pdf},
   year = {2018},
   type = {Journal Article}
}

@article{KimS2023,
   author = {Kim, Sunghwan and Das, Bamadev and Ji, Kang Hyeon and Moghaddam, Mahsa Haddadi and Chen, Cheng and Cha, Jongjin and Namgung, Seon and Lee, Dukhyung and Kim, Dai-Sik},
   title = {Defining the zerogap: cracking along the photolithographically defined Au–Cu–Au lines with sub-nanometer precision},
   journal = {Nanophotonics},
   volume = {12},
   number = {8},
   pages = {1481-1489},
   DOI = {https://doi.org/10.1515/nanoph-2022-0680},
   url = {https://doi.org/10.1515/nanoph-2022-0680},
   year = {2023},
   type = {Journal Article}
}

@article{Jelic2024,
   author = {Jelic, V. and Adams, S. and Hassan, M. and Cleland-Host, K. and Ammerman, S. E. and Cocker, T. L.},
   title = {Atomic-scale terahertz time-domain spectroscopy},
   journal = {Nature Photonics},
   volume = {18},
   number = {9},
   pages = {898-904},
   ISSN = {1749-4893},
   DOI = {10.1038/s41566-024-01467-2},
   url = {https://doi.org/10.1038/s41566-024-01467-2},
   year = {2024},
   type = {Journal Article}
}

@article{Siday2024,
   author = {Siday, T. and Hayes, J. and Schiegl, F. and Sandner, F. and Menden, P. and Bergbauer, V. and Zizlsperger, M. and Nerreter, S. and Lingl, S. and Repp, J. and Wilhelm, J. and Huber, M. A. and Gerasimenko, Y. A. and Huber, R.},
   title = {All-optical subcycle microscopy on atomic length scales},
   journal = {Nature},
   volume = {629},
   number = {8011},
   pages = {329-334},
   ISSN = {1476-4687},
   DOI = {10.1038/s41586-024-07355-7},
   url = {https://doi.org/10.1038/s41586-024-07355-7},
   year = {2024},
   type = {Journal Article}
}

@article{Park2026,
   author = {Park, Hajung and Song, Euncheol and Eom, Seonhye and Park, Hyeong-Ryeol and Hong, Sung Ju and Bahk, Young-Mi},
   title = {Ultrafast Terahertz Nanoplasmonic Platform for Subpicosecond Charge–Voltage Characterization in Metal–Insulator–Metal Nanogaps},
   journal = {ACS Photonics},
   ISSN = {2330-4022},
   url = {https://doi.org/10.1021/acsphotonics.6c01359},
   year = {2026},
   type = {Journal Article}
}

@article{Pitchappa2016,
   author = {Pitchappa, Prakash and Ho, Chong Pei and Cong, Longqing and Singh, Ranjan and Singh, Navab and Lee, Chengkuo},
   title = {Reconfigurable Digital Metamaterial for Dynamic Switching of Terahertz Anisotropy},
   journal = {Advanced Optical Materials},
   volume = {4},
   number = {3},
   pages = {391-398},
   ISSN = {2195-1071},
   DOI = {https://doi.org/10.1002/adom.201500588},
   url = {https://doi.org/10.1002/adom.201500588},
   year = {2016},
   type = {Journal Article}
}

@article{Manjappa2018,
   author = {Manjappa, Manukumara and Pitchappa, Prakash and Singh, Navab and Wang, Nan and Zheludev, Nikolay I. and Lee, Chengkuo and Singh, Ranjan},
   title = {Reconfigurable MEMS Fano metasurfaces with multiple-input–output states for logic operations at terahertz frequencies},
   journal = {Nature Communications},
   volume = {9},
   number = {1},
   pages = {4056},
   ISSN = {2041-1723},
   DOI = {10.1038/s41467-018-06360-5},
   url = {https://doi.org/10.1038/s41467-018-06360-5},
   year = {2018},
   type = {Journal Article}
}

@article{Prakash2024,
   author = {Prakash, Saurav and Pitchappa, Prakash and Agrawal, Piyush and Jani, Hariom and Zhao, Yunshan and Kumar, Abhishek and Thong, John and Linke, Jian and Ariando, Ariando and Singh, Ranjan and Venkatesan, Thirumalai},
   title = {Electromechanically Reconfigurable Terahertz Stereo Metasurfaces},
   journal = {Advanced Materials},
   volume = {36},
   number = {32},
   pages = {2402069},
   ISSN = {0935-9648},
   DOI = {https://doi.org/10.1002/adma.202402069},
   url = {https://doi.org/10.1002/adma.202402069},
   year = {2024},
   type = {Journal Article}
}

@Article{Moazzez2013,
AUTHOR = {Moazzez, Behrang and O'Brien, Stacey M. and Merschrod S., Erika F.},
TITLE = {Improved Adhesion of Gold Thin Films Evaporated on Polymer Resin: Applications for Sensing Surfaces and MEMS},
JOURNAL = {Sensors},
VOLUME = {13},
YEAR = {2013},
NUMBER = {6},
PAGES = {7021--7032},
URL = {https://www.mdpi.com/1424-8220/13/6/7021},
PubMedID = {23760086},
ISSN = {1424-8220},
}

@article{Halbritter2003,
   author = {Halbritter, A. and Csonka, Sz and Mihály, G. and Jurdik, E. and Kolesnychenko, O. Yu and Shklyarevskii, O. I. and Speller, S. and van Kempen, H.},
   title = {Transition from tunneling to direct contact in tungsten nanojunctions},
   journal = {Physical Review B},
   volume = {68},
   number = {3},
   pages = {035417},
   note = {PRB},
   DOI = {10.1103/PhysRevB.68.035417},
   url = {https://link.aps.org/doi/10.1103/PhysRevB.68.035417
https://journals.aps.org/prb/abstract/10.1103/PhysRevB.68.035417},
   year = {2003},
   type = {Journal Article}
}

@article{Jelic2017,
   author = {Jelic, Vedran and Iwaszczuk, Krzysztof and Nguyen, Peter H. and Rathje, Christopher and Hornig, Graham J. and Sharum, Haille M. and Hoffman, James R. and Freeman, Mark R. and Hegmann, Frank A.},
   title = {Ultrafast terahertz control of extreme tunnel currents through single atoms on a silicon surface},
   journal = {Nature Physics},
   volume = {13},
   number = {6},
   pages = {591-598},
   ISSN = {1745-2481},
   DOI = {10.1038/nphys4047},
   url = {https://doi.org/10.1038/nphys4047
https://www.nature.com/articles/nphys4047},
   year = {2017},
   type = {Journal Article}
}

@article{Hemmetter2021,
   author = {Hemmetter, Andreas and Yang, Xinxin and Wang, Zhenxing and Otto, Martin and Uzlu, Burkay and Andree, Marcel and Pfeiffer, Ullrich and Vorobiev, Andrei and Stake, Jan and Lemme, Max C. and Neumaier, Daniel},
   title = {Terahertz Rectennas on Flexible Substrates Based on One-Dimensional Metal–Insulator–Graphene Diodes},
   journal = {ACS Applied Electronic Materials},
   volume = {3},
   number = {9},
   pages = {3747-3753},
   ISSN = {2637-6113},
   DOI = {10.1021/acsaelm.1c00134},
   url = {https://doi.org/10.1021/acsaelm.1c00134},
   year = {2021},
   type = {Journal Article}
}

\end{document}